# Realization of a Tunable Artificial Atom at a Supercritically Charged Vacancy in Graphene

Jinhai Mao[1†], Yuhang Jiang[1†], Dean Moldovan[2], Guohong Li[1], Kenji Watanabe[3], Takashi Taniguchi[3], Massoud Ramezani Masir[2,4], Francois M. Peeters[2] and Eva Y. Andrei[1]

[1]Rutgers University, Department of Physics and Astronomy, 136 Frelinghuysen Road, Piscataway, NJ 08855 USA

[2] Departement Fysica, Universiteit Antwerpen, Groenenborgerlaan 171, B-2020 Antwerpen, Belgium

[3]Advanced Materials Laboratory, National Institute for Materials Science, 1-1 Namiki, Tsukuba 305-0044, Japan

[4]Department of Physics, University of Texas at Austin, Austin TX 78712, USA

[†] These authors contributed equally to this work.

**Graphene's remarkable electronic properties have fueled the vision of a graphene-based platform for lighter, faster and smarter electronics and computing applications. One of the challenges is to devise ways to tailor graphene's electronic properties and to control its charge carriers [1-5]. Here we show that a single atom vacancy in graphene can stably host a local charge and that this charge can be gradually built up by applying voltage pulses with the tip of a scanning tunneling microscope (STM). The response of the conduction electrons in graphene to the local charge is monitored with scanning tunneling and Landau level spectroscopy, and compared to numerical simulations. As the charge is increased, its interaction with the conduction electrons undergoes a transition into a supercritical regime [6-11] where itinerant electrons are trapped in a sequence of quasi-bound states which resemble an artificial atom. The quasi-bound electron states are detected by a strong enhancement of the density of states (DOS) within a disc centered on the vacancy site which is surrounded by halo of hole states. We further show that the quasi-bound states at the vacancy site are gate tunable and that the trapping mechanism can be turned on and off, providing a new mechanism to control and guide electrons in graphene.**

Supercriticality in atoms occurs when the Coulomb coupling, $\beta = Z\alpha$, exceeds a critical value of order unity, where Z is the atomic number and $\alpha \sim 1/137$ is the fine structure constant. In this regime the electronic orbitals, starting with the 1S state, sink into the Dirac continuum until Z is reduced to the critical value. This process, known as atomic-collapse (AC), is accompanied by vacuum polarization and the spontaneous generation of positrons [12, 13]. But accessing this new physics requires ultra-heavy nuclei which do not exist in nature. In graphene, where the effective fine structure constant, $\alpha_g = \alpha \frac{c}{v_F} \sim 2$, is much larger the critical coupling, $\beta_c = \frac{Z_c}{\kappa} \alpha_g = 0.5$, can be reached for a relatively modest charge [6-10] (c is the speed of light, $v_F$ the Fermi velocity and $\kappa$ the effective dielectric constant). The transition to the supercritical regime in graphene is marked by the emergence of a sequence of quasi-bound states which can trap electrons. However, because graphene is a good conductor it is difficult to deposit and maintain a charge on its surface. Attempts to create charge by ionizing ad-atoms or donors with an STM tip [14] showed that the charge decays when the field is removed. Alternatively, ions can be deposited directly on graphene, but because the charge transfer is inefficient [15], attaining the critical Z requires piling up many ions which, in the absence of a controllable mechanism to overcome the Coulomb repulsion, is very challenging [11].

Here we show that a vacancy in graphene can stably host a positive charge. The charge is deposited by applying voltage pulses with an STM tip and its gradual buildup is monitored with scanning tunneling spectroscopy (STS) and Landau level (LL) spectroscopy, and compared to numerical simulations. This ability to gradually charge the vacancy provides unprecedented control over the charge states allowing to explore *in situ* the evolution of the system from the subcritical to the supercritical regime. As the charge crosses into the supercritical regime we observe the sudden appearance of negative energy quasi-bound electron-states at the vacancy site, which mimic the bound states in an artificial atom. Seen as DOS peaks below the Dirac point (DP) energy, $E_D$, these states produce a high intensity disc in the STS maps which is accompanied by a circular halo of hole-states emerging at a larger radius. In addition to observing the lowest energy peak R1, which is the equivalent of the 1S atomic state reported previously [11], we observe two new states, R2 and R1'. R2 is the equivalent of the 2S atomic state and R1' is a new branch of collapse states arising from the broken sublattice symmetry at the vacancy site.

Vacancies were created by sputtering graphene with $He^+$ ions [16, 17] and probed by STM and STS in a gated sample configuration shown in Fig. 1A (methods). The STM topography in Fig. 1B shows that the initially smooth surface becomes peppered with bright defects. Zooming into a defect we observe the triangular $\sqrt{3} \times \sqrt{3}\, R30°$ interference pattern (Fig. 1B, Inset) characteristic of single Carbon vacancies [18, 19]. This structure is absent in vacancies that are passivated by trapped ions, providing a clear topographic signature that distinguishes them from bare vacancies. We here focus only on unpassivated vacancies as identified by their triangular structure in STM topography.

To evaluate the effect of a vacancy on the local DOS we measured the spatial evolution of the differential conductance (dI/dV) shown in Fig. 1C [20]. Far from the vacancy the spectrum is 'V' shaped with the minimum identifying the DP energy measured with respect to the Fermi energy, $E_F \equiv 0$. In contrast, on the vacancy site the spectrum features a sharp vacancy peak (VP) close to $E_D$, which is tightly localized within ~ 2 nm (Fig. 1C, Inset) [21]. From the doping dependence of the VP, (Fig. 1D), we find that its energy tracks the evolution of $E_D$ with gate voltage, $V_g$, indicating that it is pinned to the DP and not to $E_F$ [19, 22] (Fig. 1E). This distinction is important for understanding the evolution of the spectra with charge and for identifying the supercritical regime.

According to recent DFT calculations [23], the removal of a Carbon atom from graphene and the subsequent lattice relaxation produces a positively charged vacancy with effective charge $Z/\kappa \sim +1|e|$, where $|e|$ is the fundamental unit of charge. The effective charge and sign of a local vacancy in graphene can be directly estimated from LL spectroscopy, by measuring the on-site energy shift of the N = 0 LL relative to its value far away, $\Delta E_{00} \approx -\frac{Z}{\kappa}\frac{e^2}{4(2\pi)^{1/2}\varepsilon_0 l_B}$ [24]. Here $\varepsilon_0$ is the permittivity of free space and $l_B$ is the magnetic length. As we show in the supplementary information (SI) (Fig. 16s(b)) this expression, obtained from first order perturbation theory, is valid for $\beta > 0.1$, where it approaches the tight binding result. In Fig. 2A we show a LL map along a line-cut traversing the vacancy. Contrary to expectations we observed a negligible downshift indicating that the vacancy charge is significantly smaller than that predicted by DFT. To estimate the local charge when it is too small to produce an observable LL downshift, we compare the splitting of the N = 0 LL caused by the locally broken sublattice symmetry, to numerical simulations (section 11 in SI, Fig. 16s(a)). The best fit to the data, shown in Fig. 2C, is obtained for a positively charged vacancy with $\beta \sim 0.1$. We next attempt to increase the vacancy charge by applying STM voltage pulses at its center (methods). It is well known that such pulses can functionalize atoms, tailor the local structure or change the charge state [25, 26]. In Fig. 2B we plot the LL map after applying several pulses. Now we observe a clear downshift in the N = 0 LL as expected for a positively charged vacancy with $\beta > 0.1$. Comparing to the theoretical simulation in Fig. 2D we obtain a good match to the data for $\beta \sim 0.45$, which is also the value obtained from the formula above. The charge continues to build up as more pulses are delivered

(SI Fig. 6s). Interestingly, both positive and negative pulses produce similar positively charged states (SI Fig. 7s) and their charge is stable remaining unchanged for as long as the experiment is kept cold. These results suggest that, similar to the numerical relaxation process applied in the DFT calculations [23], pulsing helps the system relax toward the ground state which is the fully charged vacancy. The number of pulses needed to reach a given intermediate charge varies for different vacancies (Fig. 6s) consistent with the stochastic process characterizing the relaxation. Importantly, applying a similar pulse sequence to pristine graphene produces no signature of charge buildup (SI Fig. 8s). This indicates that the charge buildup occurs at the vacancy site itself and not in the substrate. Experiments were carried out on several samples and on different substrates with pulsing producing similar results (section 6 in SI).

We performed numerical simulations, solving the tight-binding Hamiltonian for a charged vacancy (section 7 in SI), to calculate the evolution of the DOS with $\beta$. The simulations consider effects which go beyond the continuum limit and Dirac equation, such as the effect of a vacancy in graphene with broken electron-hole symmetry. This allows to calculate the evolution of the local DOS well into the supercritical regime while including both the effects of the vacancy and magnetic field, which is not possible with continuum models. The results, summarized in Figs. 3A and 3B, show that at low $\beta$ in the absence of field the spectrum consists of a single peak which evolves from the uncharged VP. With increasing $\beta$ the VP broadens and its energy becomes more negative, all the while remaining tightly localized on the vacancy site (SI Fig. 5s). Upon exceeding the critical value, $\beta > 0.5$, a new branch, labeled R1, emerges below the DP. We identify R1 as the counterpart of the 1S AC state in atoms [12,13]. R1 is clearly distinguishable from the VP by its significantly larger spatial extent (SI Figs. 11s and 15s) which reflects its quasi-bound nature [8,9]. With increasing $\beta$, R1 develops a satellite, R1', which tracks its evolution with β. While R1 is a universal feature of AC states, R1' is due to the locally broken sublattice symmetry and is peculiar to the supercritically charged single-atom vacancy. As β further increases more branches emerge, starting with R2 which is the equivalent of the 2S AC state [7-9].

In Fig. 3 we consider the evolution of the spectra with β. After the first few pulses the peak, which initially is close to the DP, broadens and sinks in energy eventually disappearing below the experimental horizon. This behavior together with the fact that the peak remains tightly localized on the vacancy site (SI Fig. 5s) helps identify this peak with the VP branch in the simulation. To obtain the vacancy charge we map the peak energy after each pulse onto the simulated VP in Fig. 3A (symbols in figure). This gives the value of β in the subcritical regime. As the charge is further increased, new peaks emerge below the DP which are qualitatively different from the VP: they are very sensitive to gating and are significantly more spatially spread out, suggesting that they belong to the R1 branch. Mapping their energies onto the R1 branch we obtain the β values ranging from $\beta = 0.95$ to $1.29$, indicating that the charge is supercritical. Comparing the simulated (Fig. 3B) and measured (Fig. 3C) spectra we find that the AC states gradually emerge with increasing charge, first revealing the R1 state, then its satellite R1' and finally for $\beta=1.29$ all three states R1, R1' and R2 are clearly resolved.

The spatial dependence of the spectra in the AC regime (Fig. 4) shows that, in contrast to the tightly localized VP, the AC states extend far beyond the vacancy site. The DOS enhancement associated with the R1 state (top right of Fig. 4C) consists of a central disk surrounded by an outer halo. Comparing to the local DP (dashed line) in Fig. 4B, we note that within the central disc R1 lies above the local DP corresponding to electron states, but it is below the local DP at the halo position, indicating spatially separated electron-hole states. The hole states are absent at energies far from R1 as well as in the subcritical regime which excludes standard energy resolved Friedel oscillations [8] [section 10 in SI]. An alternative interpretation of the electron-hole states is suggested by the analogy with atomic collapse in superheavy nuclei, where vacuum polarization generates electron-positron pairs [7, 8, 13, 27].

The doping dependence of the AC state shown in Fig. 5 provides insight into its screening by the itinerant carriers. We find that the screening displays strong electron-hole asymmetry. In the p-doped regime the R1 peak persists over the entire experimental range, suggesting poor screening by the positive carriers. In contrast, on the n-doped side the AC peaks disappear rapidly with doping indicating that screening by negative carriers is very efficient. This unusual screening asymmetry may reflect the charge induced electron-hole asymmetry in the local DOS (Fig. 5 and SI Fig. 10s) or it could be the result of correlations [11], but more work is needed to understand this phenomenon. A direct experimental consequence of the asymmetry is that the AC state can be turned on or off with a modest gate voltage, providing a switching mechanism for localized states. Moreover, the readily implemented charging technique reported here could enable the fabrication of artificial atom arrays for electrostatic control and guidance of electrons in graphene.

**Methods**

The samples consist of two stacked graphene layers deposited on a hexagonal boron nitride (hBN) flake (G/G/BN) which rests on a $SiO_2$ substrate capping an n-doped Si backgate, Fig. 1A. The hBN and bottom graphene layers reduce the substrate-induced random potential fluctuations and help reveal the intrinsic electronic properties of the top layer [28-30]. The atomic lattices of the two graphene layers are intentionally misaligned in order to decouple them and to restore the linear spectrum characteristic of single layer graphene [31]. The decoupling is directly observable with Landau Level (LL) spectroscopy (SI Fig. 1s and Fig. 2s) through the characteristic $\sqrt{NB}$ dependence of the LL energies expected for single layer graphene [20, 32]. Here, B is the magnetic field and N the level index. Vacancies were created by sputtering with 100-140 eV He$^+$ ions under UHV conditions followed by high temperature *in situ* annealing [16, 17]. The effect of the irradiation can be seen in the Raman spectra by the appearance of the D peak signifying the presence of lattice vacancies (Fig. 4s). The STM voltage pulses are typically 2-3V and ~1-10 sec duration. STM spectra are taken with the tip grounded and the bias voltage, $V_b$, applied to the sample. Multiple vacancies were studied in two samples which are labeled 1 (Fig. 1) and 2 (Figs. 2-5) and their characteristics and gate voltage dependence are described in the SI. Furthermore, our studies of pulsed vacancies in graphene supported by other substrates (G/BN, G/G/SiO$_2$) produced similar results as discussed in Section 6 of the SI.

**Acknewledgements**

Funding was provided by DOE-FG02-99ER45742 (STM/STS), NSF DMR 1207108 (fabrication and characterization). Theoretical work supported by ESF-EUROCORES-EuroGRAPHENE, FWO-VI and Methusalem program of the Flemish government. We thank V. F. Libisch, M. Pereira and E. Rossi for useful discussions.

**Author contributions**

J.M. and Y.J. collected and analyzed data and wrote the paper; G.L built the STM and analyzed the data; D.M., M.R.M. and F.M.P. carried out the theoretical work; K.W. and T.T. provided the BN sample. E.Y.A. directed the project, analyzed the data and wrote the paper.

**Additional information**

The authors declare no competing financial interests. Supplementary information accompanies this paper on www.nature.com/naturephysics. Reprints and permission information is available online at http://npg.nature.com/reprintsandpermissions. Correspondence and requests for materials should be addressed to E.Y.A.

# Figure captions:

Figure 1. **Electronic properties of single vacancy on G/G/BN.** (**A**) Schematics of experimental setup. (**B**) Typical STM topography of G/G/BN surface before and after He$^+$ ion sputtering. Inset: atomic resolution topography of a 4nm × 4nm area containing a single vacancy shows the triangular interference pattern characteristic of a bare (unpassivated) vacancy. (**C**) Spatial dependence of the vacancy peak (VP). Curves are vertically shifted for clarity. Inset is the dI/dV map at the energy of the VP. Experimental parameters for panels (B) and (C): $V_b =$ −300mV, I = 20pA, $V_g = -30$V. (**D**) Doping dependence of the VP. The arrows label the Dirac point energy, $E_D$, of pristine graphene 50nm away from any vacancy. Curves are offset for clarity. Tunneling parameters: $V_b = -200$mV, I = 20pA. (**E**) Gate voltage dependence of $E_D$ (black) and VP energy (blue) relative to the Fermi energy ($E_F$=0) extracted from the dI/dV spectra. The dotted line is a fit as detailed in section 2 SI.

Figure 2. **Effect of charged vacancy on the Landau level (LL) spectra.** Spatial dependence of the LLs across a line-cut traversing the vacancy before (A) and after (B) applying the voltage pulse. The position of the vacancy is marked by the enhanced intensity of the VP residing at its center. The energy shift of the N=0 LL at 6T, $\Delta E_{00}$ ~ 45 mV, corresponds to an effective charge of Z/κ ~ 0.2 and β ~ 0.45 indicating that the charge is subcritical. We note that the N=0 LL is split at the vacancy site. Since the wavefunction for the N=0 LL is confined to one sublattice, the removal of one atom breaks the local symmetry and produces the strong splitting. (C), (D) Numerical simulation of spatial dependence of the LL spectra across a line-cut reproduces the experimental results in (A) and (B) for β ~ 0.1 and β ~ 0.45 respectively.

Figure 3. **Electronic properties of the charged vacancy.** (A) Simulated map of the evolution of the spectra with β. The intensity scale in the VP regime is divided by 2 to facilitate the comparison with the atomic collapse (AC) regime. The symbols represent the energies of the VP and AC peaks from panel C. (B) Simulated spectra for the β values in (C). Curves are vertically offset for clarity. (C) Evolution of STS with charge (increasing from bottom to top). Each curve is marked with its corresponding β value. The horizontal dashed line separates between spectra in the subcritical and supercritical regimes. The vertical dashed line represents

the bulk Dirac point (DP) measured far from any vacancy. Experimental parameters: $V_b = -200$ mV, I = 20 pA, $V_g = -54$ V.

Figure 4. **Spatial evolution of the AC state**. (A) Spatial dependence of the AC states for $\beta = 1.29$. The dashed line labels the bulk DP. (B) Spatial dependence of simulated spectra for $\beta = 1.29$. In the simulation, the bulk DP is taken as the energy origin. The dashed curve represents the local DP determined from the Coulomb potential (U) of the charged vacancy with $\beta = 1.29$, $U(r) = \beta \dfrac{\hbar v_F}{r}$. (C) Constant energy dI/dV maps in the vicinity of the vacancy. Top left: map of neutral vacancy taken at $-91$meV. The other panels represent maps of the charged vacancy in (A) taken at the energies corresponding to R1 ($-91$meV top right), R1' ($-25$meV, bottom left) and R2 (50meV, bottom right). Tunneling parameters $V_b = -200$ mV, I = 20 pA, $V_g = -54$ V.

Figure 5. **Gate dependent screening of AC states.** (A) Carrier density dependence of the AC state. Arrows label the bulk DP measured far from the vacancy. The horizontal dashed line separates the p-doped (below) from the n-doped regime (above) and the vertical dashed line marks $E_F$. The spectra are labeled by the carrier density expressed in units of $10^{11}$ cm$^{-2}$. The carrier density is obtained from the dI/dV curves far from the vacancy: $= \dfrac{1}{\pi}\left(\dfrac{E_D}{\hbar v_F}\right)^2$, $v_F$ is extracted from the LL fitting. Tunneling parameters: $V_b = -100$ mV, I = 20 pA. (B) Evolution of $\beta$ with carrier density shows the strong electron-hole asymmetry of the screening. The square marks the inflection point at the crossing of the AC with $E_F$.

References:

1. M. I. Katsnelson, K. S. Novoselov, A. K. Geim, Chiral tunnelling and the Klein paradox in graphene. *Nat Phys* **2**, 620 (2006).
2. N. M. R. Peres, F. Guinea, A. H. Castro Neto, Electronic properties of disordered two-dimensional carbon. *Physical Review B* **73**, 125411 (2006).
3. T. Ando, Screening Effect and Impurity Scattering in Monolayer Graphene. *Journal of the Physical Society of Japan* **75**, 074716 (2006/07/15, 2006).
4. S. Das Sarma, S. Adam, E. H. Hwang, E. Rossi, Electronic transport in two-dimensional graphene. *Reviews of Modern Physics* **83**, 407 (2011).
5. V. N. Kotov, B. Uchoa, V. M. Pereira, F. Guinea, A. H. Castro Neto, Electron-Electron Interactions in Graphene: Current Status and Perspectives. *Reviews of Modern Physics* **84**, 1067 (2012).
6. V. R. Khalilov, C.-L. Ho, Dirac electron in a Coulomb field in (2+1) dimensions. *Modern Physics Letters A* **13**, 615 (1998).
7. V. M. Pereira, J. Nilsson, A. H. Castro Neto, Coulomb Impurity Problem in Graphene. *Physical Review Letters* **99**, 166802 (2007).
8. A. V. Shytov, M. I. Katsnelson, L. S. Levitov, Vacuum Polarization and Screening of Supercritical Impurities in Graphene. *Physical Review Letters* **99**, 236801 (2007).
9. A. V. Shytov, M. I. Katsnelson, L. S. Levitov, Atomic collapse and quasi–Rydberg states in graphene. *Physical Review Letters* **99**, 246802 (2007).
10. M. M. Fogler, D. S. Novikov, B. I. Shklovskii, Screening of a hypercritical charge in graphene. *Physical Review B* **76**, 233402 (2007).
11. Y. Wang *et al.*, Observing Atomic Collapse Resonances in Artificial Nuclei on Graphene. *Science* **340**, 734 (May 10, 2013, 2013).
12. I. Pomeranchuk, Y. Smorodinsky, About the energy levels of systems with Z >1/137. *J. Fiz. USSR* **9**, 97 (1945).
13. Y. B. Zeldovich, V. S. Popov, Electronic structure of superheavy atoms. *Soviet Physics Uspekhi* **14**, 673 (1972).
14. Y. Wang *et al.*, Mapping Dirac quasiparticles near a single Coulomb impurity on graphene. *Nat Phys* **8**, 653 (2012).
15. C. E. Junkermeier, D. Solenov, T. L. Reinecke, Adsorption of NH2 on Graphene in the Presence of Defects and Adsorbates. *The Journal of Physical Chemistry C* **117**, 2793 (2013/02/14, 2013).
16. O. Lehtinen *et al.*, Effects of ion bombardment on a two-dimensional target: Atomistic simulations of graphene irradiation. *Physical Review B* **81**, 153401 (2010).
17. J.-H. Chen, L. Li, W. G. Cullen, E. D. Williams, M. S. Fuhrer, Tunable Kondo effect in graphene with defects. *Nat Phys* **7**, 535 (2011).
18. K. F. Kelly, D. Sarkar, G. D. Hale, S. J. Oldenburg, N. J. Halas, Threefold Electron Scattering on Graphite Observed with C60-Adsorbed STM Tips. *Science* **273**, 1371 (September 6, 1996, 1996).
19. M. M. Ugeda, I. Brihuega, F. Guinea, J. M. Gómez-Rodríguez, Missing Atom as a Source of Carbon Magnetism. *Physical Review Letters* **104**, 096804 (2010).
20. E. Y. Andrei, G. Li, X. Du, Electronic properties of graphene: a perspective from scanning tunneling microscopy and magneto-transport. *Reports on Progress in Physics* **75** 056501 (2012).
21. O. V. Yazyev, L. Helm, Defect-induced magnetism in graphene. *Physical Review B* **75**, 125408 (2007).
22. V. M. Pereira, F. Guinea, J. M. B. Lopes dos Santos, N. M. R. Peres, A. H. Castro Neto, Disorder Induced Localized States in Graphene. *Physical Review Letters* **96**, 036801 (2006).

23. Y. Liu, M. Weinert, L. Li, Determining charge state of graphene vacancy by noncontact atomic force microscopy and first-principles calculations. *Nanotechnology* **26**, 035702 (2015).
24. A. Luican-Mayer *et al.*, Screening Charged Impurities and Lifting the Orbital Degeneracy in Graphene by Populating Landau Levels. *Physical Review Letters* **112**, 036804 (2014).
25. J. Repp, G. Meyer, F. E. Olsson, M. Persson, Controlling the Charge State of Individual Gold Adatoms. *Science* **305**, 493 (July 23, 2004, 2004).
26. A. Zhao *et al.*, Controlling the Kondo Effect of an Adsorbed Magnetic Ion Through Its Chemical Bonding. *Science* **309**, 1542 (September 2, 2005, 2005).
27. Y. Nishida, Vacuum polarization of graphene with a supercritical Coulomb impurity: Low-energy universality and discrete scale invariance. *Phys Rev B* **90**, 165414 (2014).
28. C. R. Dean *et al.*, Boron nitride substrates for high-quality graphene electronics. *Nature Nanotechnology* **5**, 722 (Oct, 2010).
29. J. Xue *et al.*, Scanning tunnelling microscopy and spectroscopy of ultra-flat graphene on hexagonal boron nitride. *Nature Materials* **10**, 282 (2011).
30. C. P. Lu *et al.*, Local and Global Screening Properties of Graphene Revealed through Landau Level Spectroscopy. *arXiv:1504.07540*, (2015).
31. G. Li *et al.*, Observation of Van Hove singularities in twisted graphene layers. *Nature Physics* **6**, 109 (Feb, 2010).
32. G. Li, A. Luican, E. Y. Andrei, Scanning Tunneling Spectroscopy of Graphene on Graphite. *Physical Review Letters* **102**, 176804 (2009).

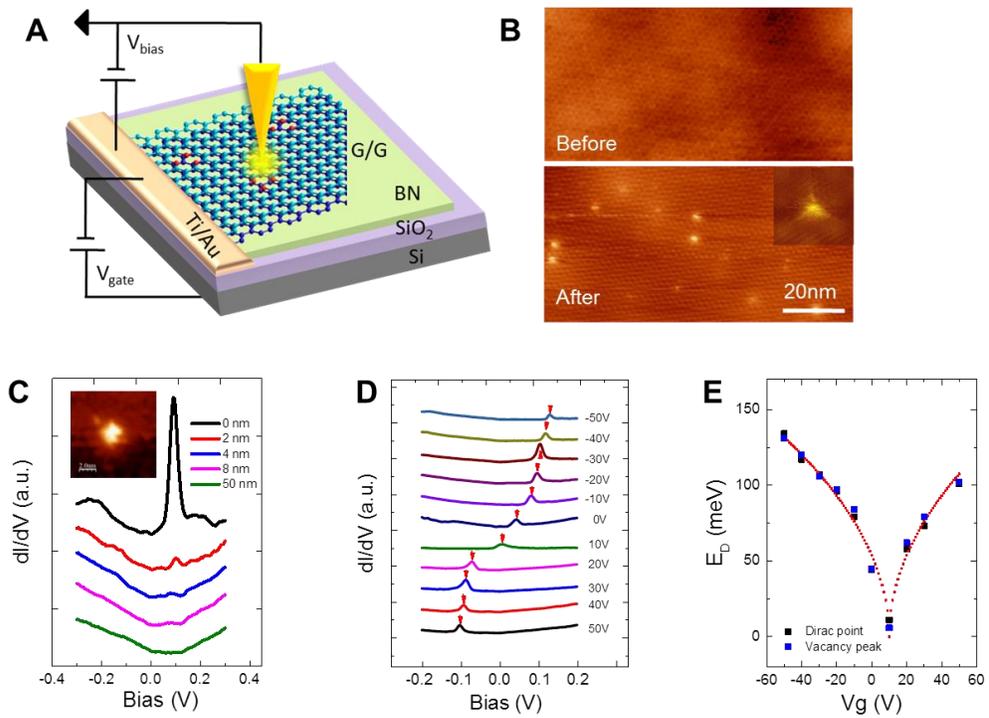

Figure 1

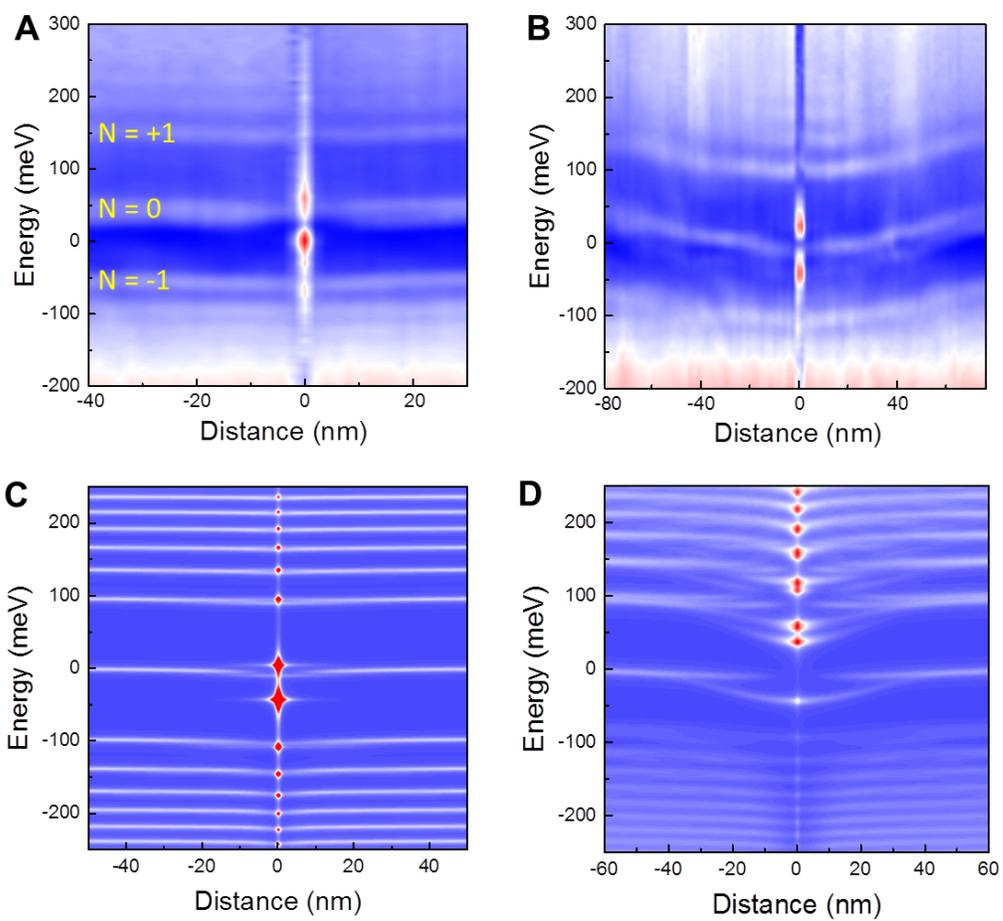

Figure 2

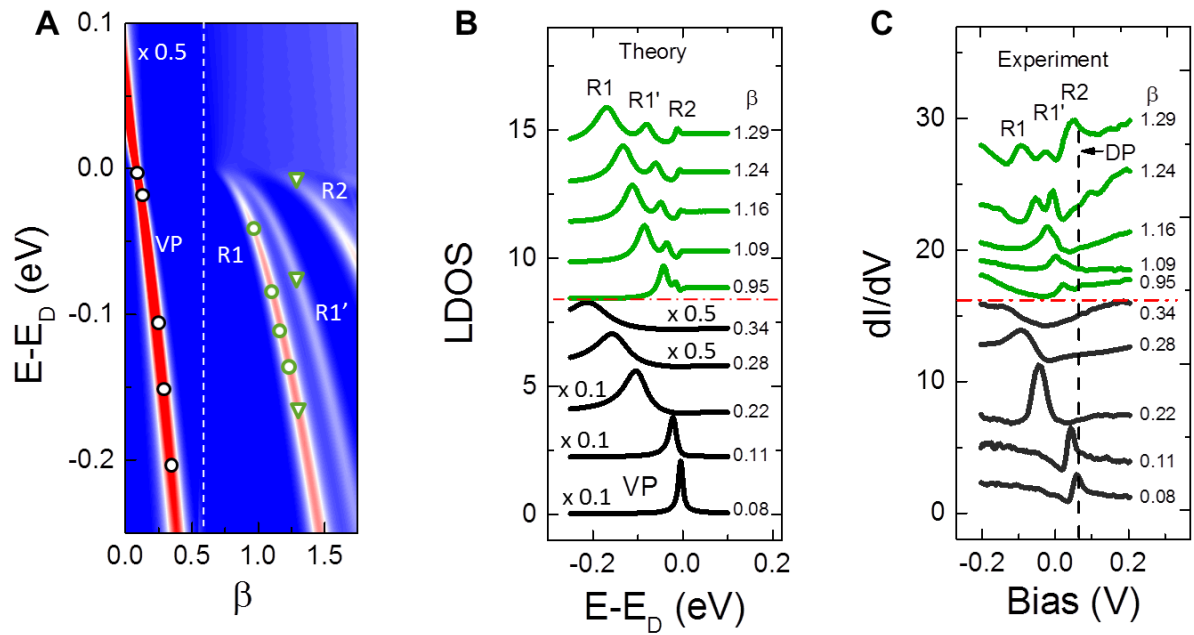

Figure 3

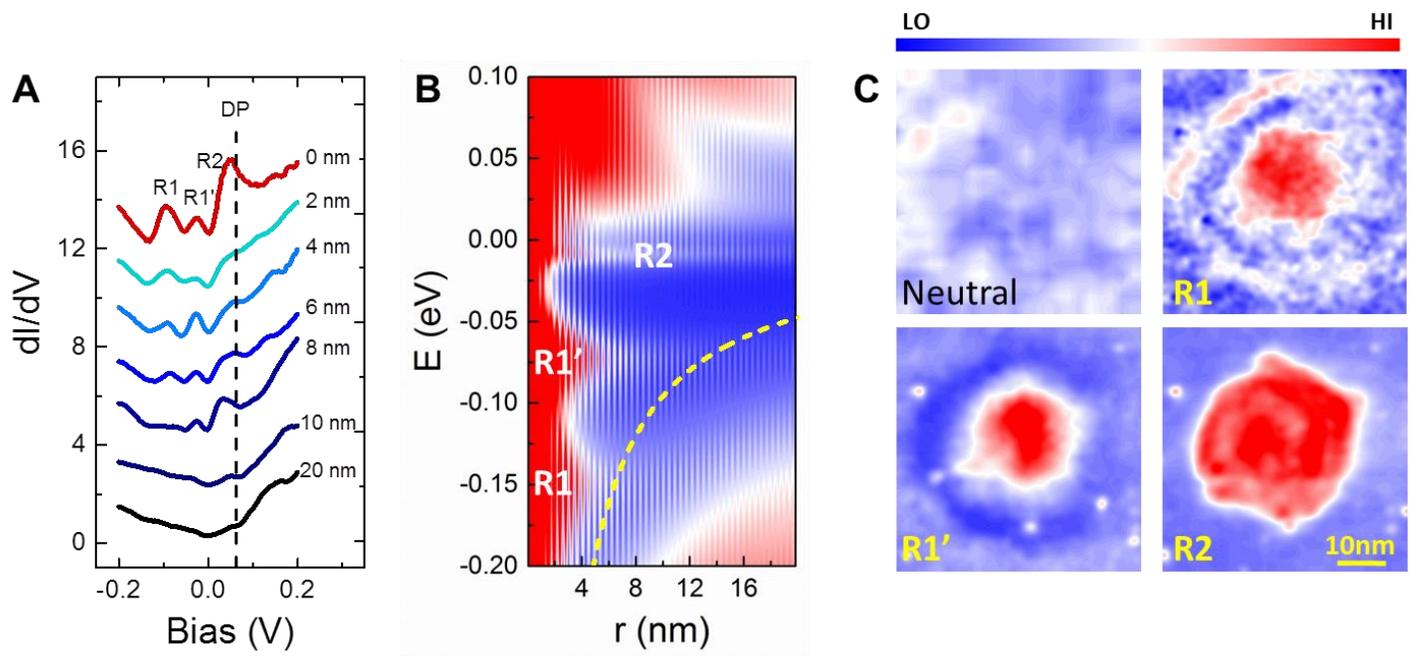

Figure 4

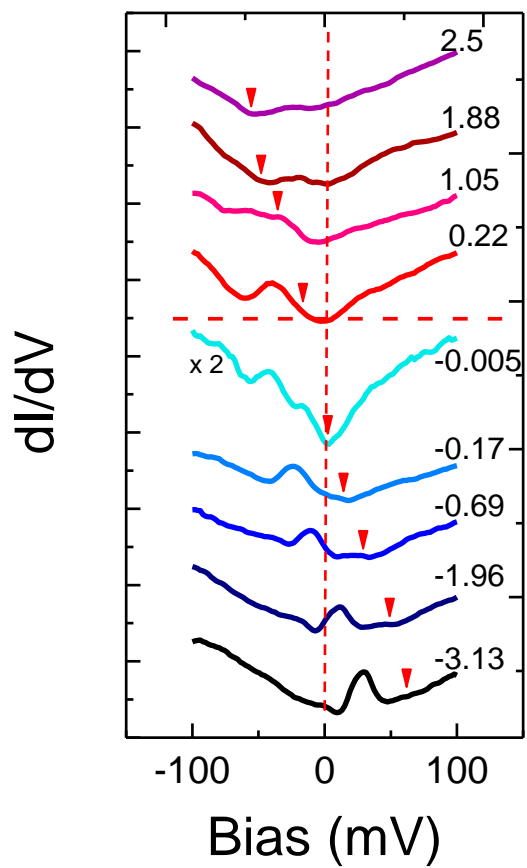 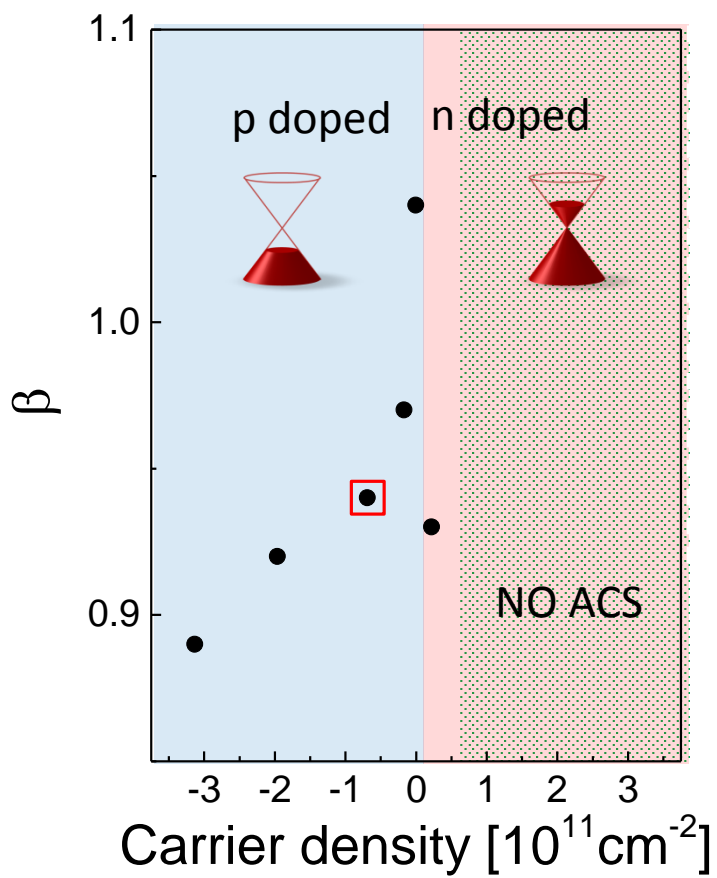

Figure 5

# Supplementary Information

## Realization of a Tunable Artificial Atom at a Charged Vacancy in Graphene

Jinhai Mao[1], Yuhang Jiang[1], Dean Moldovan[2], Guohong Li[1], Kenji Watanabe[3], T. Taniguchi[3], Massoud Ramezani Masir[2,4], Francois M. Peeters[2] and Eva Y. Andrei[1]

[1]Rutgers University, Department of Physics and Astronomy, Piscataway, NJ 08855 USA

[2]Departement Fysica, Universiteit Antwerpen, B-2020 Antwerpen, Belgium

[3]Advanced Materials Laboratory, NIMS, Tsukuba 305-0044, Japan

[4]Department of Physics, University of Texas at Austin, Austin TX 78712, USA

## 1. Sample fabrication and characterization

i) *Sample fabrication*. In this work, we use G/G/BN, G/BN and G/G on $SiO_2$ to perform the experiment. For the G/G/BN sample, hBN thin flakes were exfoliated onto the $SiO_2$ surface. Subsequently the first graphene layer was deposited on the hBN flake via a dry transfer process using a sacrificial PMMA thin film. Before stacking the top layer graphene, the PMMA was removed with acetone and IPA, followed by furnace annealing in forming gas (10% $H_2$ and 90% Ar) at 230 ℃ for 3 hrs. The second layer graphene was stacked by using the same procedure as the first layer. Au/Ti electrodes were deposited by the standard SEM lithography for the STM contact. After the liftoff process, the sample was annealed again in furnace with forming gas to remove the PMMA residues. Subsequently, the sample is loaded in the UHV chamber for further annealing at 230 ℃ overnight. To generate the single vacancies, the sample is exposed to a 100 -140 eV $He^+$ ion beam followed by high temperature *in situ* annealing. The other stacked samples, $G/BN/SiO_2$ and $G/G/SiO_2$, were prepared by the same procedure.

ii) *Characterization by STM topography and Landau level (LL) spectroscopy*. The STM experiment is performed at 4K using a cut PtIr tip. The dI/dV spectroscopy is performed using the standard lockin method [1,2] with bias modulation typically 2mV at 473.1Hz. To charge the single vacancies, voltage pulses are applied directly at the desired vacancy site with the STM tip at ground potential.

The intrinsic electronic properties of graphene can be effectively isolated from the random potential induced by the SiO$_2$ substrate by using an intermediate graphene layer and an hBN buffer underneath the layers [3,4]. When the twist angle between the two stacked graphene layers exceeds 10° the two layers are electronically decoupled at the experimentally relevant energies. Therefore a large twist angle was chosen to ensure a linear dispersion near the Dirac point. We use two G/G/BN samples during this work. For sample 1, the atomic resolution STM topography on the top graphene layer in Fig. 1s (a) shows the honeycomb structure which is a signature of the electronic decoupling from the bottom layer. LL spectroscopy provides a direct way to prove the layer decoupling [3]. For single layer graphene, the energy level sequence is given by: $E_N = sgn(N)v_F\sqrt{2e\hbar|N|B}$, where N is the LL index, $v_F$ is the Fermi velocity, $\hbar$ is the reduced Plank constant. For fixed magnetic field, the root N dependence for the sequence is the fingerprint of single layer graphene. Fig.1s (b) shows the LL spectrum of the G/G/BN sample at 6T. By fitting the LLs sequence (Fig. 1s (c)), we obtain the value of the Fermi velocity $v_F = (0.95 \pm 0.03) \times 10^6 \, m/s$.

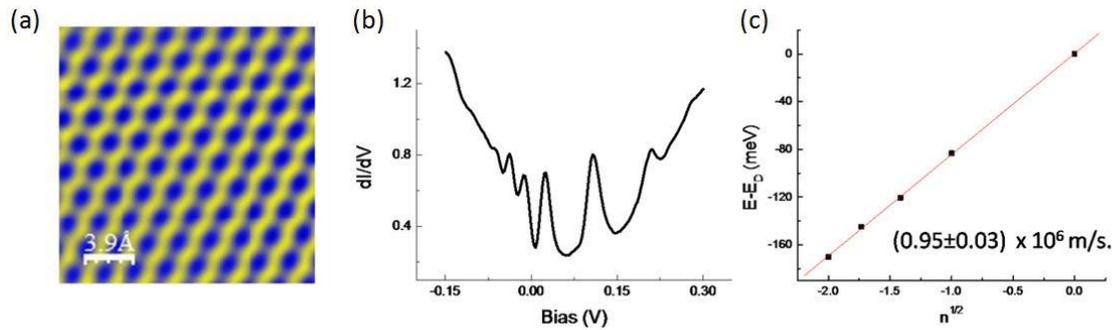

**Figure 1s.** Electronic properties of decoupled twisted bilayer graphene (Sample 1). (a) Atomic resolution of the top graphene layer, $V_b$ = -300mV, I = 20pA. (b) dI/dV curve on graphene at B=6T with $V_b$ = -300mV and I = 20pA. (c) Fit of the Landau level sequence in (b) used to extract the Fermi velocity.

Similar to sample 1, we plot out the LLs (Fig. 2s (a)) for sample 2. By fitting the LLs sequences for this sample (Fig. 2s (b)), we obtain $v_F = (0.98 \pm 0.01) \times 10^6 \, m/s$.

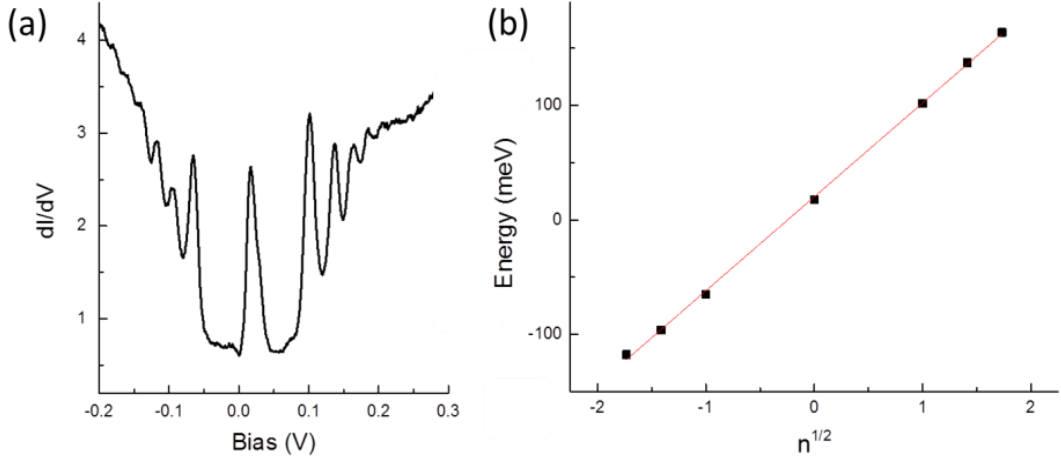

**Figure. 2s** Electronic properties of decoupled twisted bilayer graphene (Sample 2). (a) dI/dV curve on graphene at B=6T with $V_b$ = -300mV and I = 20pA. (b) Fit of the Landau level sequence in (a) used to extract the Fermi velocity

## 2. Gate dependence of Dirac point for the top layer graphene

The bulk Dirac points were obtained from the backgate dependence of the dI/dV curves near the vacancy before applying the voltage pulses at the vacancy site. This is used as the reference point for calculating the beta values for the charged impurity. Fig. 1D (Sample 1) and Fig.3s (a) (Sample 2) shows the dI/dV curves with arrows labeling the Dirac point for each backgate voltage.

The total charge on the twisted double-layer graphene can be estimated from the parallel plate capacitor model. However, as recent transport measurement have shown, there may be a charge imbalance between the two layers owing to the partial screening from the bottom layer. As a result, the charge in the top layer (the one accessed experimentally) does not necessarily correspond to that calculated from the parallel plate capacitor model. In this case the formula which assumes an equal distribution of charge between the layers: $E_D = \hbar v_F \sqrt{\alpha \pi |V_g - V_D|/2}$, is not valid ($\alpha$ is capacitance per area, $V_D$ gate voltage offset of the neutrality point and 2 represents the layer degeneracy). There are many examples in the literature of twisted double layer graphene where the movement of $E_D$ in the top layer differs significantly from that of the bottom layer [5-10].

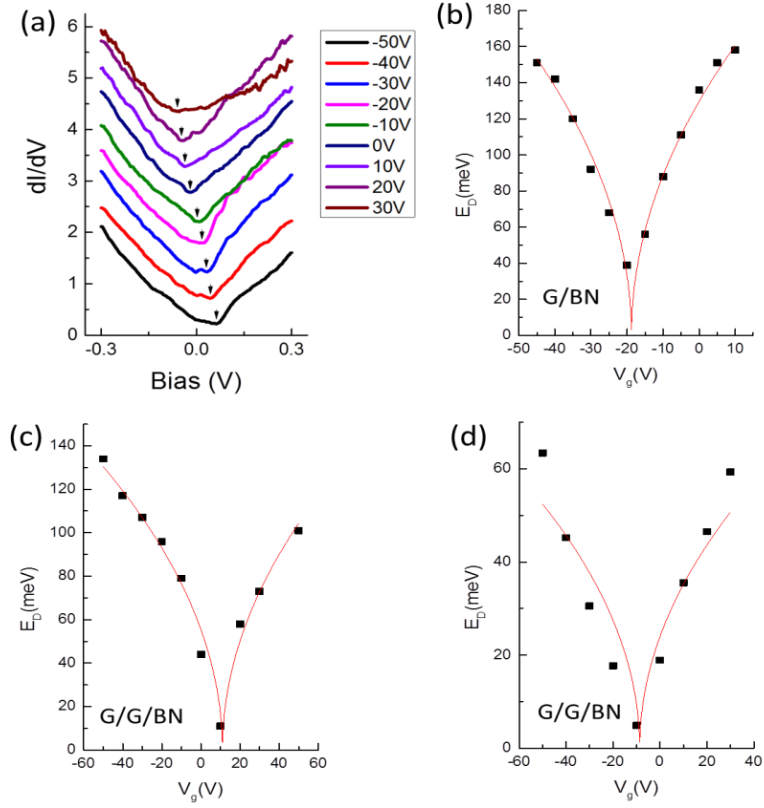

**Figure 3s** Gating effect for top-layer graphene in different samples. (a) Backgate dependence dI/dV curves to determine Dirac point for Sample 2 in the maintext, $V_b$=-300 mV, and I = 20pA. (b)-(d) Fitting of Dirac point with backgate voltage for G/BN, G/G/BN (sample 1) and G/G/BN (sample 2) respectively.

To gain insight into the effect of gating on the carrier density in the top layer we fit the Dirac point movement in the top layer to the expression $E_D = \hbar v_F \sqrt{\nu \alpha \pi |V_g - V_D|}$ where $\alpha = 7.0 \times 10^{10} cm^{-2} V^{-1}$ and $\nu \equiv \frac{n_{top}}{n_{total}}$ is the ratio of the carrier density in the top graphene layer ($n_{top}$) to the total carrier density ($n_{total}$). Using the values of $v_F$ obtained from LL spectroscopy to fit the data in Fig. 3s for the different samples we find ν ~1 for G/BN where $v_F = 1.06 \times 10^6 m/s$, ν ~0.3 for G/G/BN sample 1 where $v_F = 0.95 \times 10^6 m/s$, and ν ~0.13 for sample 2 where $v_F = 0.98 \times 10^6 m/s$. We note that for the single layer graphene, the value ν ~1 is in good agreement with the result expected from the parallel plate capacitor model. For a double layer with evenly distributed charge in the top and bottom layers one would expect $\nu = 0.5$. The values $\nu < 0.5$ obtained in our twisted double-layer samples indicates that the top layer is less populated than the bottom one. This model provides a simple empirical way to estimate the distribution of the total charge between the two layers.

The charge distribution between the layers depends on the interlayer dielectric environment or interlayer capacitance which varies widely from sample to sample. Values ranging between $0.6 \mu F cm^{-2}$ to $6.9 \mu F cm^{-2}$ have been reported (Refs. 5 and 6), with the variations being attributed to differences in interlayer distance, Fermi velocity, interlayer coupling and coupling to substrate. The effect of a constant interlayer capacitance, $C_{int}$, on the Dirac point movement in a uniform twisted double-layer graphene is captured by the following equations: [6]

$$e(V_{BG} - V_D) = \frac{e^2(n_T + n_B)}{C_{BG}} + E_F(n_B) \quad (1)$$

$$E_F(n_B) = \frac{e^2 n_T}{C_{int}} + E_F(n_T) \quad (2)$$

Substituting (2) in (1) and using the relation $E_F = \hbar v_F \sqrt{\pi n}$, to fit the data for the two twisted graphene samples studied here, we find $C_{int}$ values of $4.2 \mu F cm^{-2}$ (Sample 1) and $0.5 \mu F cm^{-2}$ (Sample 2) respectively.

## 3. Effect of irradiation characterized by Raman spectroscopy

To confirm the sputtering efficiency, we carried out Raman spectroscopy on a control sample of CVD graphene on $SiO_2$, using the same bombardment parameters. The Raman spectra before and after the sputtering differ by the appearance of a pronounced D peak, shown in Fig. 4s, which is a direct consequence of the short range scattering introduced by the vacancies.

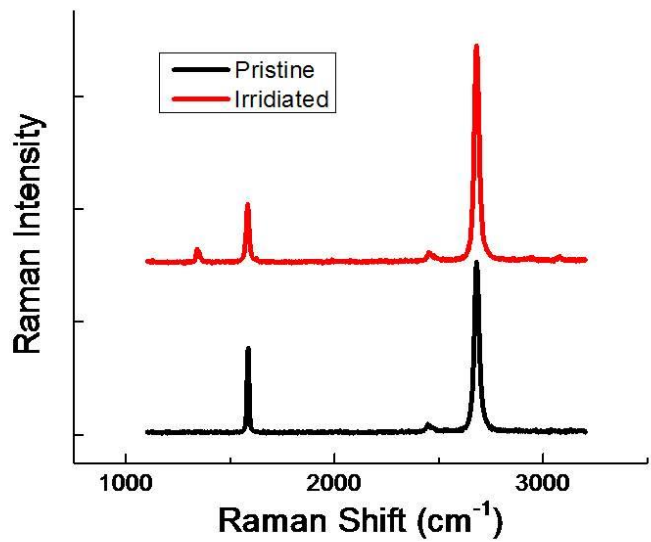

**Figure 4s.** Comparison of Raman spectra on the clean and irradiated graphene samples.

## 4. Spatial dependence of vacancy peak in the subcritical regime.

Applying the voltage pulse through the tunneling junction causes a buildup of positive charge on the single vacancy site. As reported in the main text, in the subcritical regime the vacancy peak moves deeper below the bulk Dirac point with increasing charge. In this regime the peak in the DOS remains tightly localized, within 2nm, on the vacancy site, similar to the behavior of the neutral vacancy peak shown in Fig. 1C of the main text. Fig. 5s shows the spatial dependence of the vacancy peak in the subcritical regime.

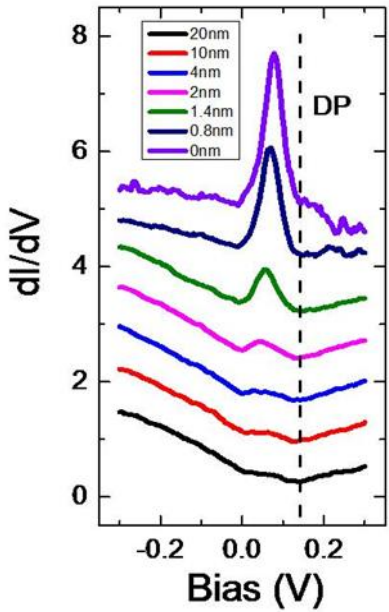

**Figure 5s.** Spatial dependence of the vacancy peak in the subcritical regime. The black dashed line is the bulk Dirac point measured far from any vacancy site. The STM parameters are $V_b = -300$ mV and $I = 20$ pA.

## 5. Building up charge by pulsing vacancies

In Fig. 6s we show the evolution of the onsite charge with pulsing for 8 different vacancies. We find that the charge increases with the administration of pulses in an irreversible way. Furthermore, we found that the charge accumulated on the vacancy does not depend on the pulse polarity. In Fig. 7s we show the evolution of the spectra following positive voltage pulses. Similarly to the case of negative voltage pulses, the vacancy peak energy shift down with every consecutive positive pulse. For the positive pulses we found that a threshold voltage of 2V to induce an observable charging effect. These results suggest that the pulses help the system relax towards the theoretically predicted charged ground state, as discussed in the main text.

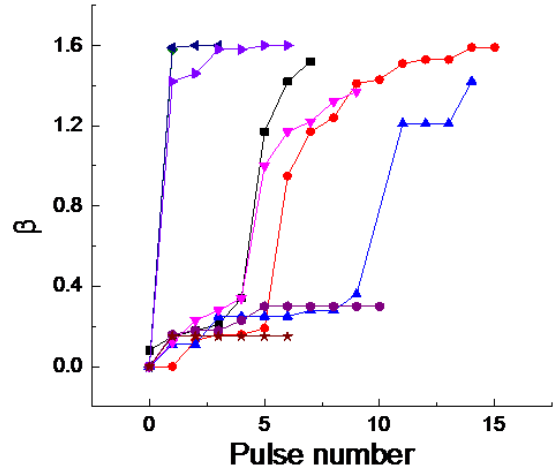

**Figure 6s.** Evolution of on-site charge with the pulsing sequence for eight vacancies. The pulse polarity was negative – with the tip grounded and the negative voltage pulse applied to the sample. The β values are determined by mapping the vacancy peak energy onto the simulated branch for β < 0.5 and the atomic collapse (AC) peak energy onto the respective simulated branches, R1, R1' or R2.

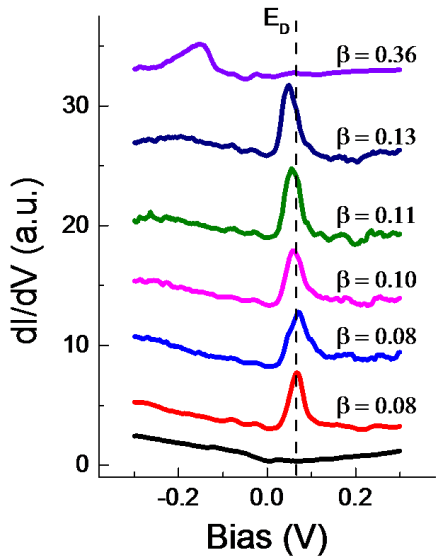

**Figure 7s.** Evolution of the vacancy peak with charge buildup following the application of positive pulses. The vacancy peak moves monotonically to lower energy below the Dirac point energy $E_D$. The black curve is taken far from the vacancy and the red curve at the vacancy site before the first pulse.

## 6. Pulse effect on pristine graphene and irradiated graphene on different substrates

### a. Pulse effect on pristine graphene

In order to examine the possibility that the charge accumulates in the substrate rather than at the vacancy site we applied similar pulse sequences to the pristine parts of the G/G/BN surface far from a vacancy and followed the evolution of the spectra. The results are summarized in Fig. 8s. We observe no measurable shift of the local Dirac point to indicate the presence of local charging, even after applying twelve pulses. In contrast, the same sequence of pulses applied to a vacancy site resulted in significant charge buildup. This leads to the conclusion that the charge is hosted by the vacancy and not in the hBN substrate.

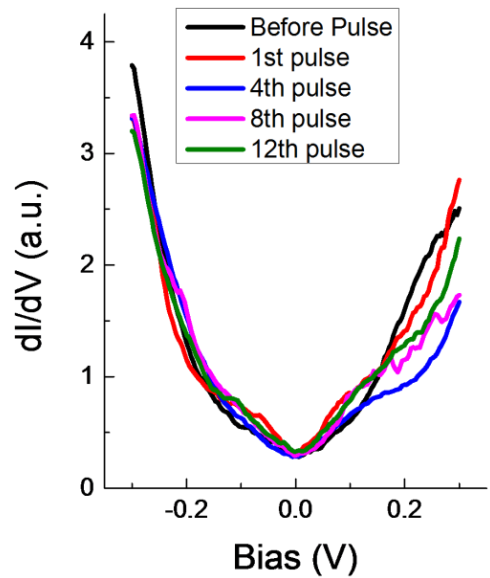

**Figure 8s.** Spectra taken on pristine graphene after administering voltage pulses show that local Dirac point does not shift within the measurement resolution. Tunneling parameters: $V_b$ = -300 mV, I = 20 pA.

### b. Pulse effect on vacancy on G/G/SiO$_2$

We further explored the role of the substrate by pulsing a vacancy site in a double graphene layer deposited on SiO$_2$, G/G/SiO$_2$ (Fig. 9s). As before, we first observed the vacancy peak at the Dirac point, and then applied the voltage pulse at the vacancy site. As with the case of the vacancy in G/G/BN, we observed a shift of the vacancy peak to lower energy after applying pulses. Comparing the results on G/G/BN and G/G/SiO$_2$ surface, we conclude that the substrate does not play a crucial role in the charging process.

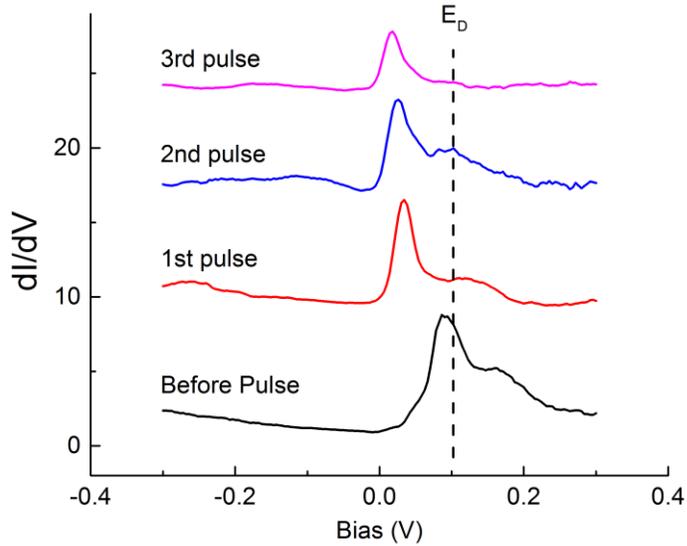

**Figure 9s.** Charging a single vacancy in G/G/SiO$_2$ by the pulse method. After each voltage pulse (-2V for 1s), the vacancy peak shifts further below E$_D$, as is the case for the charge vacancy in the G/G/BN/SiO$_2$ sample discussed in the main text.

### c. Pulse effect on vacancy on G/BN

To address the question of the role of the bottom layer graphene in the charging process we applied the same pulsing sequence to vacancies in a single layer graphene deposited on hBN, Similar to the case of the vacancy in G/G/BN discussed in the main text, applying voltage pulses to a vacancy in G/BN also charges it into the supercritical regime, as illustrated in Fig. 10s. Also similar to the double layer case, we find that screening of the positive charge exhibits a strong electron-hole asymmetry shown in Fig. 10s (c). In Fig. 11s we show the spatial dependence of the atomic collapse state (ACS). Unlike the vacancy peak which is tightly localized the ACS peak is far more extended. In addition we note that the spectra exhibit strong electron hole asymmetry, indicative of the positive charge.

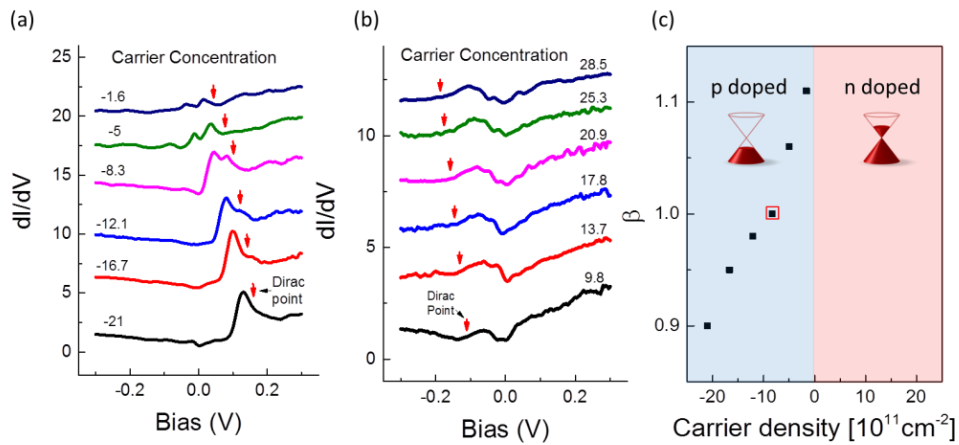

**Figure 10s.** (a, b) Carrier concentration dependence of AC state of charged vacancy in a G/BN sample in units of $10^{11}$ cm$^{-2}$. Similar to the AC state in G/G/BN, a single vacancy in G/BN can also be charged into the supercritical regime by applying voltage pulses with the STM tip. Red arrows label the Dirac point obtained from the dI/dV curves far from the vacancy, at each gate voltage. (c) Carrier density dependence of the effective charge shows a strong electron-hole screening asymmetry, similar to the case of the G/G/BN sample discussed in the main text. The strong electron-hole asymmetry of the spectra provides additional evidence that the vacancy is positively charged.

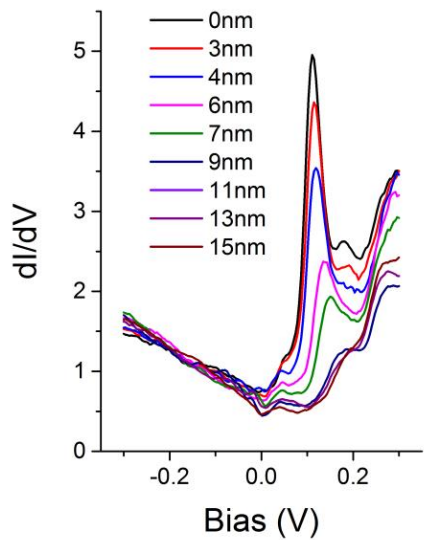

**Figure 11s.** Spatial dependence of the dI/dV curves of the supercritically charged vacancy in G/BN. The large spatial extent of the AC state stands in strong contrast to the tightly localized (~2nm) vacancy peak. Tunneling parameters: $V_b$ = -300mV, I = 20pA, Vg = -45V.

## 7. Theoretical simulation for the charged vacancies in graphene

The theoretical simulation is based on the tight-binding model (up to second nearest neighbor hopping) where the LDOS is calculated using the Chebyshev polynomial method [11]. A vacancy with $N_v$ missing atoms is modeled by removing the corresponding real space elements from the tight-binding Hamiltonian matrix. The charge is hosted by the vacancy in the graphene plane. To account for the finite size of the charged vacancy a truncated potential is used:

$$V(r) = \begin{cases} -\hbar v_F \frac{\beta}{r_0}, & \text{if } r \leq r_0 \\ -\hbar v_F \frac{\beta}{r}, & \text{if } r > r_0 \end{cases}, \quad (1)$$

where $r_0$ is the radius of the constant potential region at the center of the charge, and $\beta$ is the dimensionless coupling constant defined in the main text. The solutions close to the Dirac point are fairly insensitive to the detailed form of the central region $r \leq r_0$. The effect of the choice of $r_0$ is discussed below. Unless otherwise stated the simulations are done with $r_0 = 0.5$ nm.

Close to $\beta = 0$ we can observe the vacancy as a peak in Fig. 3B of the main text. The position of the vacancy peak decreases rapidly in energy with increasing $\beta$ and by the time R1 appears it is well out of the measurement range (for $\beta = 1$ it appears around -1 eV). Collapse resonance R1 splits into two peaks which behave similarly as a function of $\beta$. The R1 is a feature of the collapse, independent of the vacancy, as shown in Fig. 12s. The R1' peak is unique to the single-atom vacancy.

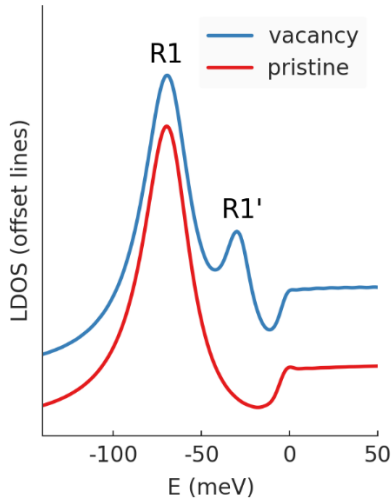

**Figure 12s.** Simulated LDOS for a charge of $\beta = 1.05$ in pristine graphene and the same charge hosted by a single-atom vacancy. The LDOS lines are offset for clarity.

## 8. r₀ influence on the AC state resonance peak

The form of the truncated potential, Eq. (1), takes the finite size of the charge into account via the parameter $r_0$. Based on experimental data, we set this parameter to $r_0 = 0.5$ nm. In order to investigate the effect of this parameter we varied it as shown in Fig. 13s (a). Reducing the size of $r_0$ increases the strength of the potential and in the limit $r_0 \to 0$ it corresponds to a point charge. As we move closer to this limit, the energy levels R1 and R1' diverge: R1 dives to very low energy while increasing its intensity, but R1' stays close to the Dirac point while losing intensity. R1 and R1' are mostly made up of opposite sublattice states. As shown in Fig. 13s (b), sublattice A is closer to the center with a radius of $a_{cc} = 0.142$ nm (carbon-carbon distance), while the minimum radius of sublattice B is farther away at $a = 0.246$ nm (lattice constant). This defines the minimum radii of the R1 and R1' resonances. Fig. 13s (c) shows a zoomed in view at small $r_0$ and the vertical lines correspond to the radii from Fig. 13s (b). Resonance R1' has a larger radius which prevents it from diving to low energy in the potential of Eq. (1). It loses most of its intensity in the region where $a_{cc} < r_0 < a$. Resonance R1 significantly increases its intensity in the region $r_0 < a$ and saturates at $a_{cc}$. In the region $r_0 < a_{cc}$, the potential radius becomes smaller than the vacancy and is therefore non-physical.

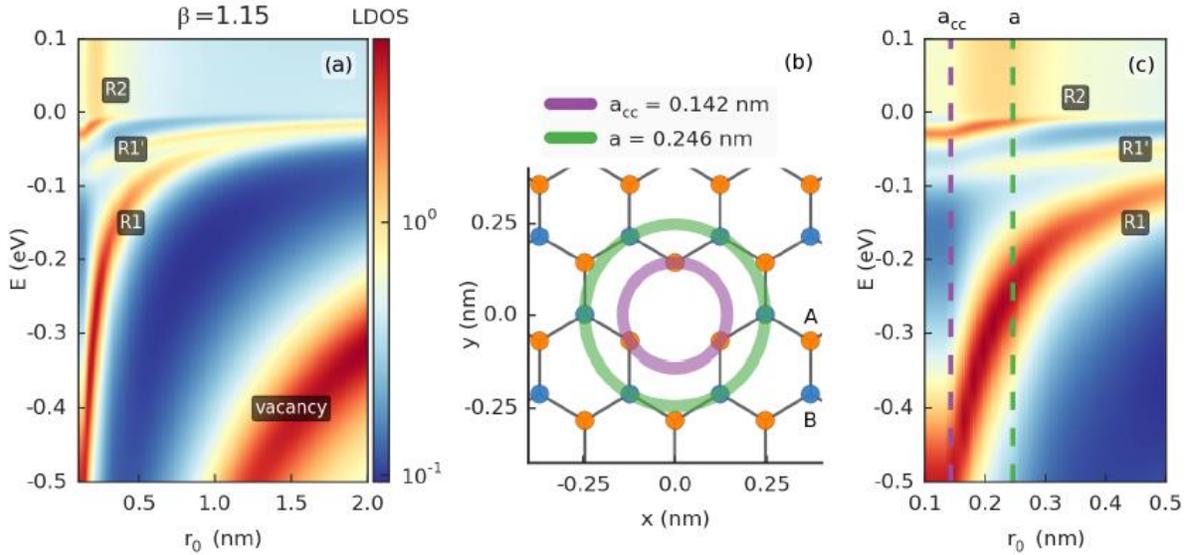

**Figure 13s.** (a) LDOS as function of the $r_0$ parameter for $\beta = 1.15$. (b) Illustration of the model system around the vacancy. The circles indicate minimum radii of sublattices A and B. (c) Zoomed in view of panel (a) Close to the center. The vertical lines correspond to the radii from panel (b).

## 9. Lattice symmetry of the vacancy state

A vacancy may consist of $N_v = N_A + N_B$ missing carbon atoms, where $N_A$ and $N_B$ correspond to the number of atoms removed from sublattices A and B respectively. In the presence of electron-hole symmetry, introducing a vacancy with $N_A \neq N_B$ creates a zero energy state which is quasilocalized near the vacancy. In addition, this state exists only on a single sublattice, corresponding to the one with the lower number of removed atoms [12].

We investigate neutral vacancies in graphene. Fig. 14s illustrates four types of vacancies labeled 1 to 4 according to the number of missing carbon atoms. The LDOS is shown for each vacancy as well as for pristine graphene (0 missing atoms). Vacancies with 1, 3 and 4 missing atoms exhibit a LDOS with a very high intensity peak at the Dirac point which corresponds to the quasilocalized zero energy state. The vacancy with 2 missing atoms retains a V shaped LDOS with only a slightly increased slope compared to pristine graphene. This is because it preserves the sublattice symmetry, with one missing atom from each sublattice. The other vacancies have a sublattice difference $N_B - N_A$ of 1, 1 and -2 for vacancies 1, 3 and 4, respectively.

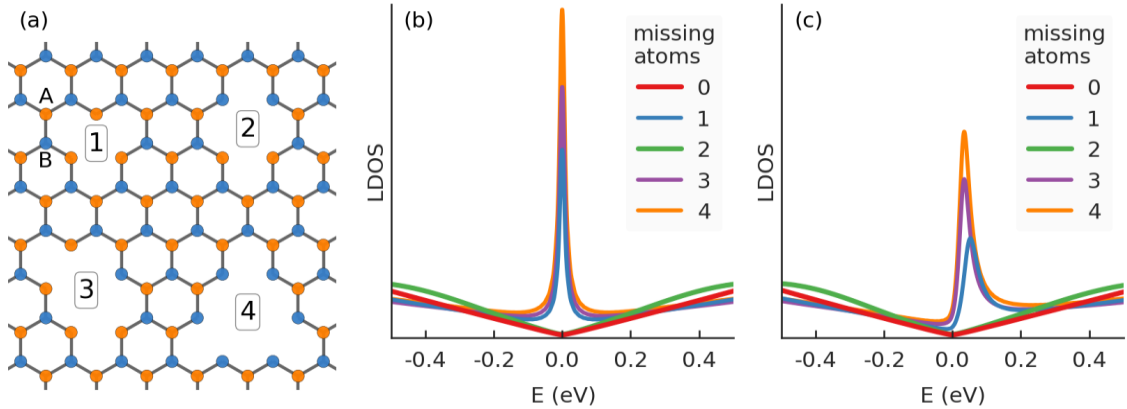

**Figure 14s.** (a) Illustration of vacancies with 1 to 4 missing atoms. (b) Simulated LDOS at 1 nm from the center of each type of vacancy including only the first nearest neighbor hopping. (c) The same results when the next nearest hopping ($t' = 0.1$ eV) is also included.

The vacancy peak in Fig. 14s (b) is symmetric because it is calculated with just the first nearest neighbor tight-binding model. Adding the next-nearest neighbor hopping term ($t' \neq 0$) produces electron-hole asymmetry. In this case the zero energy state turns into a resonance for which the linewidth and the displacement (from zero energy) are both proportional to the electron-hole asymmetry (i.e. the value of the $t'$ parameter). In contrast to the vacancy state, we found that the next-nearest neighbor term influences very weakly the AC states.

## 10. Spatial extent of the atomic collapse states

In Fig. 15s we compare the experimental and theoretical plots of the spatial dependence of the R1, R1' and R2 AC states with that of the uncharged vacancy state. For the R1 and R1' maps both experiment and theory show a high intensity central region (red) which corresponds to quasibound electron states. After that, there is a low intensity ring (blue) which is the barrier, followed by a slightly higher intensity region (white) which is the hole continuum. The R2 state has a larger radius so the outer hole region is outside the limits of the figure. The vacancy state is quite different. It only has the high-intensity center, but no corresponding outer hole region.

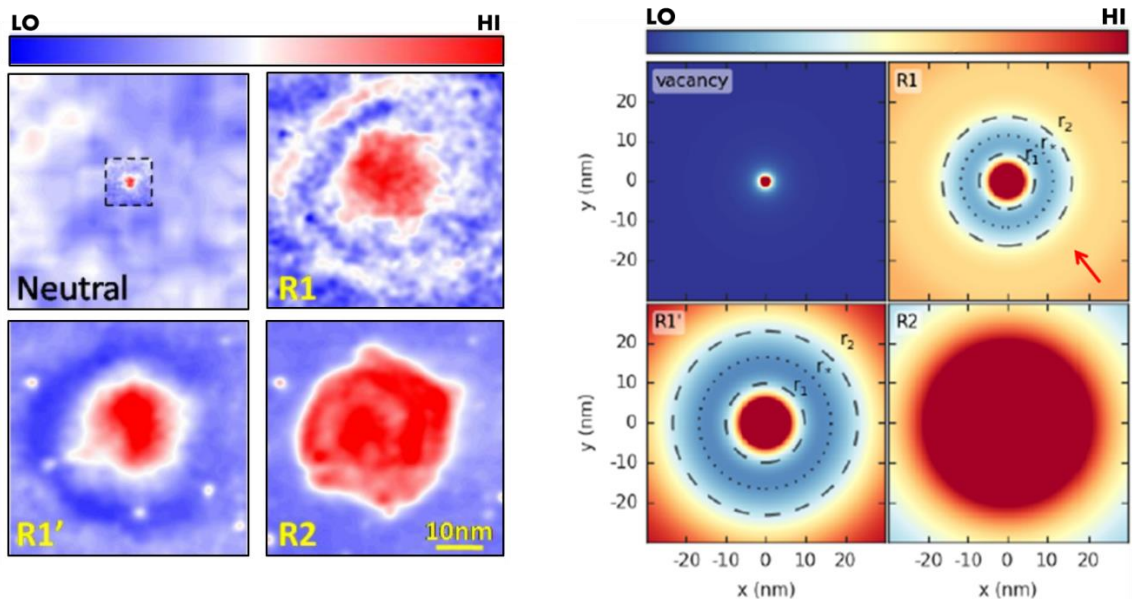

**Figure 15s.** (Left) Experimental constant energy dI/dV maps in the vicinity of the vacancy, as in Fig. 4 C in the main text. The central dashed square in the top left panel represents data taken with higher spatial resolution in order to resolve the VP. (Right) Simulated spatial map of the LDOS for the vacancy at β = 0 and E = 0 and R1, R1', R2 at β = 1.25 and E = −0.15, −0.07 and −0.01 eV, respectively.

## 11. Landau level bending in the presence of a charged vacancy

The local sublattice symmetry breaking of the vacancy creates two additional levels above and below the N = 0 LL, marked VL+ and VL− in Fig. 16s where the simulated data is shown. The experiment and simulation match well as shown in Fig. 2 of the main text. Fig. 2 (B and D) of the main text shows the spatial dependence with β = 0.45. In this case the N = 0 LL level shifts down by $E_{sim} \approx 43$ meV in the simulation, compared to $E_{exp} \approx 45$ meV in the experiment.

Fig. 16s (a) shows the LLs as a function of β. The lower vacancy level (VL−) quickly dives to negative energy as β is increased (just like the vacancy peak in the zero-field case). The positive level (VL+) merges with the N = 0 LL as β increases. In Fig. 16s (b) we compare the β dependence of the zero LL shift obtained from our tight-binding simulation to the perturbation theory expression in Ref. 3:

$$\Delta E_{00} = -\frac{Z}{\kappa} \frac{e^2}{4(2\pi)^{1/2} \epsilon_0 l_B}. \quad (2)$$

This expression was derived for a free charge without a vacancy. The two results differ at low β where the vacancy perturbs the N=0 LL (as can be seen in Fig.16s (a)). With increasing $\beta$ the simulated $\Delta E_{00}$ curve approaches that in Eq. (2) and crosser over to a linear β dependence for $\beta > 0.12$. At the highest values of β the difference between the two lines is about 3mV.

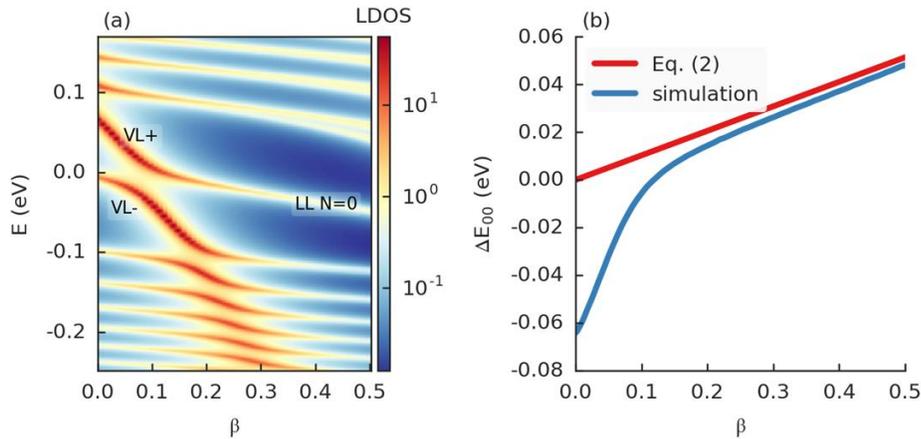

**Figure 16s** (a) Landau level LDOS as a function of β at the position of the vacancy. (b) The simulated shift of the N = 0 LL as a function of β (blue line) is compared to Eq. (2) (red line).

# References

1. Luican, A., Li, G. & Andrei, E. Y. Scanning tunneling microscopy and spectroscopy of graphene layers on graphite. *Solid State Commun.* **149**, 1151-1156 (2009).
2. Andrei, E. Y., Li, G. & Du, X. Electronic properties of graphene: a perspective from scanning tunneling microscopy and magneto-transport. *Rep. Prog. Phys.* **75**, 056501 (2012).
3. Luican-Mayer, A. *et al.* Screening Charged Impurities and Lifting the Orbital Degeneracy in Graphene by Populating Landau Levels. *Phys. Rev. Lett.* **112**, 036804 (2014).
4. Lu, C.-P. *et al.* Local and Global Screening Properties of Graphene Revealed through Landau Level Spectroscopy. *arxiv1504.07540* (2015).
5. Schmidt, H. *et al.* Tunable graphene system with two decoupled monolayers. *Appl. Phys. Lett.* **93**, 172108 (2008).
6. Fallahazad, B. *et al.* Quantum Hall effect in Bernal stacked and twisted bilayer graphene grown on Cu by chemical vapor deposition. *Phys. Rev. B* **85**, 201408 (2012).
7. Kim, S. *et al.* Coulomb drag of massless fermions in graphene. *Phys. Rev. B* **83**, 161401 (2011).
8. Sanchez-Yamagishi, J. D. *et al.* Quantum Hall Effect, Screening, and Layer-Polarized Insulating States in Twisted Bilayer Graphene. *Phys. Rev. Lett.* **108**, 076601 (2012).
9. Kim, S. *et al.* Direct Measurement of the Fermi Energy in Graphene Using a Double-Layer Heterostructure. *Phys. Rev. Lett.* **108**, 116404 (2012).
10. Beechem, T. E., Ohta, T., Diaconescu, B. & Robinson, J. T. Rotational Disorder in Twisted Bilayer Graphene. *ACS Nano* **8**, 1655-1663 (2014).
11. Covaci, L., Peeters, F. & Berciu, M. Efficient Numerical Approach to Inhomogeneous Superconductivity: The Chebyshev-Bogoliubov-de Gennes Method. *Phys. Rev. Lett.* **105**, 167006 (2010).
12. Pereira, V. M., Lopes dos Santos, J. M. B. & Castro Neto, A. H. Modeling disorder in graphene. *Phys. Rev. B* **77**, 115109 (2008).